\documentstyle[12pt,epsfig,picinpar]{article}
\def\lsim{\:\raisebox{-0.5ex}{$\stackrel{\textstyle<}{\sim}$}\:}
\def\gsim{\:\raisebox{-0.5ex}{$\stackrel{\textstyle>}{\sim}$}\:}

\begin{document}
 
\title {SUPERSYMMETRY AT THE ELECTROWEAK SCALE.
\thanks{Supported in part by the Polish Committee for Scientific
        Research and European Union Under contract CHRX-CT92-0004.}}
\author{Piotr H. Chankowski and Stefan Pokorski\\
Institute of Theoretical Physics, Warsaw University\\
ul. Ho\.za 69, 00--681 Warsaw, Poland.}
 
\maketitle
 
\vspace{-12cm}
\begin{flushright}
{\bf hep-ph/9607326} \\
\end{flushright}
\vspace{12cm}
  
\begin{abstract}
The simplest interpretation of the global success of the Standard
Model is that new physics decouples well above the electroweak scale.
Supersymmetric extension of the Standard Model offers the possibility
of light chargino and the right-handed stop (with masses below $M_Z$),
and still maintaining the successful predictions of the Standard Model.
The value of $R_b$ can then be enhanced up to $\sim 0.218$ (the Standard Model
value is $\sim 0.216$). Light chargino and stop give important contribution
to rare processes such as $b\rightarrow s \gamma$,  $\overline K^0-K^0$ and
$\overline B^0-B^0$ mixing but consistency 
with experimental results is maintained
in a large region of the parameter space. The exotic four-jet events reported
by ALEPH (if confirmed)  may constitute a signal for supersymmetry with such
a light spectrum and with explicitly broken $R-$parity. Their  interpretation
as pair production of charginos with $m_C\sim 60$ GeV, with subsequent
decay $C\rightarrow \tilde t_R b \rightarrow dsb$ (where $m_{\tilde t}\sim
55$ GeV) leads to signatures very close to the experimental observations.
\end{abstract}


{\bf 1. Introduction.} 
\vskip 0.3cm

The Standard Model (SM) is a higly successful theory. The simplest
interpretation of this fact is that any new physics decouples well above
the electroweak scale. On the other side, there are the well
known theoretical reasons (hierachy problem, "naturalness" of elementary
scalars etc.) to expect new physics to be close to the electroweak scale.
Is this in contradiction with the success of the SM? Sypersymmetry is a
particularly attractive extention of the SM. We are facing several interesting
questions such as: how low the scale of supersymmetric particle masses
can be to maintain consistency with the success of the SM? is there a room
for some superpartners with masses at the electroweak scale? or, maybe
there are even some hints for such a light spectrum?
Recent important progress in the low energy supersymmetric phenomenology
allows to address those questions in fully quantitative way, with conclusions
which are very stimulating for further experimental search for supersymmetry
at the LEP200 and the Tevatron!

The success of the SM is best measured by its description of the bulk of
the electroweak data, with an excellent overall agreement. The only clear
discrepancy is the present experimental value of ~$R_b=0.2211\pm 0.0016$
which is more than $3\sigma$ away from the theoretical prediction. 
The experimental value of ~$R_c$ ~is ~
$1.6\sigma$ ~away from the prediction and this is statistically much 
less significant. Finally, there are two ~$\sim2\sigma$ ~deviations in the
leptonic left-right asymmetry and the parameter ~${\cal A}_{b}$. ~Both
measurements come from SLAC and those deviations look merely like experimental 
problems of some mismatch between the SLAC and LEP data. Indeed, ~$A_{LR}^e$ ~
is a measure of the ~$\sin^2\theta^{eff}_{lept}$ ~and it disagrees with 
the LEP measurements of this angle which can be extracted from the parameters ~
${\cal A}_{e}$, ~${\cal A}_{\tau}$, ~$A_{FB}^l$. ~The direct SLAC measurement
of ~${\cal A}_{b}$ ~disagrees with the indirect LEP determination from ~
$A_{FB}^b$. ~

One remark is in order here. The SLD value of ~$A_{LR}^e$ ~gives ~
$\sin^2\theta^{eff}_{lept}=0.23049\pm0.00050$ ~whereas the LEP value is ~
$\sin^2\theta^{eff}_{lept}=0.23178\pm0.00031$ ~\cite{LEPEWWG}. 
In the SM, the value 
of ~$\sin^2\theta^{eff}_{lept}$ ~can be very precisely calculated (instead 
of being determined from a global fit) in terms of ~$M_Z$, ~
$m_t$ ~and ~$M_h$. ~We get e.g. the results shown in Table 1.

\begin{table*}[hbt]
\caption{Predictions in the SM for $\sin^2\theta^{eff}_{lept}$
for various top quark and Higgs boson masses. The error of this
predictions (coming mainly from the uncertainty of the hadronic contribution 
to the photon vacuum polarization) is ~$\pm0.00025$.}
\label{tab:sinus}
\begin{tabular}{|l||l|l|l|l|l|l|l|l|l|} \hline
$m_t$                         & 170   & 180   & 190 
                              & 170   & 180   & 190     \\ \hline
$M_h$                         &  60   &  60   &  60 
                              & 150   & 150   & 150     \\ \hline
$\sin^2\theta^{eff}_{lept}$   &0.23135&0.23101&0.23066
                              &0.23182&0.23149&0.23114\\ \hline
\end{tabular}
\end{table*}

It is clear that the SLD measurement favours 
larger (smaller) values of ~$m_t$ ~
($M_h$) ~($m_t\approx (180-190)$ GeV, ~$M_h\approx60$ GeV) ~
which are larger (smaller) than those favoured by the LEP result for ~
$\sin^2\theta^{eff}$ ~($m_t\approx 170$ GeV, ~$M_h\approx150$ GeV). ~
Such values give, however, worse fit to other electroweak
observables ~($M_W, \Gamma_Z, \sigma_h, A^{0,b}_{FB}$).

The precision of the data is already high enough to be sensitive to the Higgs
boson mass (which enters into the calculations only logarithmically). The 
full SM fit (the fitted parameters are ~$m_t$, ~$M_h$ ~and ~
$\alpha_s(M_Z)$) gives ~$M_h=76^{+93}_{-44}(1\sigma)^{+277}_{-76}(2\sigma)$. ~
whereas in the fit without ~$R_b$ ~and ~$R_c$ ~we get ~
$M_h=94^{+117}_{-55}(1\sigma)^{+346}_{-94}(2\sigma)$. 
We observe that the fitted value of ~$M_h$ ~does not depend much on 
whether ~$R_b$ ~is included or not into the fit. This is important in view
of the large deviation in ~$R_b$. However, some caution in the conclusions
is still necessary: if both ~$R_b$ ~and ~$A_{LR}^e$ ~are absent from the 
fitted observables we get ~
$M_h=205^{+226}_{-116}(1\sigma)^{+660}_{-170}(2\sigma)$. ~Thus, the data are 
consistent with a light Higgs boson but the ~$2\sigma$ ~upper bound depends 
strongly on the inclusion of the SLD result for ~$A_{LR}^e$ ~in the fit.

Another point of recent interest is the value of ~$\alpha_s(M_Z)$ ~
obtained from the electroweak fits
We get ~$\alpha_s(M_Z)=0.122\pm0.005$ ~and this value is somewhat larger 
then the value  obtained from the deep inelastic scattering data \cite{LOW} ~
$\alpha_s(M_Z)=0.112\pm0.005$.

We can interpret the SM fits as the MSSM fits with all
superpartners heavy enough to be decoupled. 
Explicit calculations show that this occurs for superpartner
masses already in the range 300-500 GeV. Supersymmetry then
provides a rationale for a light Higgs boson: $M_h \sim {\cal O}$(100 GeV)
and,  since the best fit in the SM is consistent with the Higgs boson mass 
precisely in this range, 
the MSSM with heavy enough superpartners gives
as good a fit to the precision electroweak data as the SM.
This is the first interesting conclusion: the success of the SM
is perfectly consistent with the supersymmetric Higgs sector
and with the soft supersymmetry breaking scale in the range
which maintains theoretical motivation for supersymmetry.
\vskip 0.5cm

{\bf 2. The  $R_b$  anomaly.}
\vskip 0.3cm

The MSSM with all superpartners heavy enough to decouple at the
electroweak scale faces the same problem as the SM, namely of the $R_b$
anomaly. The measurement of $R_b$ is difficult and the possibility
of larger than estimated systematic errors 
should be kept in mind. Still, the present
result can at least be taken as a statistical hint towards a value
of ~$R_b$ ~somewhat larger than the SM prediction.
It is then an interesting question if supersymmetry can enhance the 
value of  ~$R_b$. ~The issue has been addressed in a number of papers
\cite{RBC_S,LANER,KANE,MY_MSSM,JA_BRU,KAWE,ELONA,MY_RB}. 
It is well known already for some 
time that in the MSSM there are new contributions to the ~$Z^0\overline bb$ ~
vertex which can indeed significantly enhance the value of ~$R_b$ ~(but do 
not change ~$R_c$) ~if some superpartners are sufficiently light 
\cite{BF,RBC_S,KANE,MY_MSSM,JA_BRU,KAWE,ELONA}. More specifically, 
for low (large) ~$\tan\beta$ ~the dominant contributions are chargino--stop
($CP-$odd Higgs boson and chargino--stop) loops. 
Moreover it is also known that
new physics in ~$\Gamma_{Z^0\rightarrow\bar bb}$ ~and therefore
additional contribution to the total hadronic width of the ~$Z^0$ ~
boson would lower the fitted value of ~
$\alpha_s(M_Z)$ ~\cite{BV,LANER,SHIF}, in better agreement with its
determination from low energy data \cite{LOW}.

Any improvement in  ~$R_b$ ~must maintain the perfect agreement of the SM 
with the other precision LEP measurements and must be consistent with 
several other experimental constraints (which will be listed later on).
The bulk of the precision data, such as ~$M_W$, ~$\Gamma_Z$, ~
$\sin^2\theta^{eff}_{lept}$ ...,~
are mostly sensitive to the ~$\Delta\rho$ ~parameter which measures 
the violation of the custodial ~$SU_V(2)$ ~symmetry. The contribution
of the top$-$bottom quark  mass spliting to ~$\Delta\rho$ ~leaves very little
room for new contributions: ~$\Delta\rho<0.0015$ ~at ~95\% C.L.
\cite{LANGACKER}.
Therefore, in order to maintain the overall good agreement of the fit
with the data we must avoid new sources of the 
custodial ~$SU_V(2)$ ~symmetry breaking in the left currents.
This is assured if the left squarks of the third
generation and all left sleptons are  sufficiently heavy
\cite{MY_DR,MY_MSSM}, say, $>{\cal O}(300~{\rm GeV})$.  
At the same time, an increase in ~$R_b$ ~is sensitive 
mainly to the masses and couplings of
the  right handed top squark, charginos and - in the
case of large  $\tan\beta$ - of  the right  handed sbottom and ~
$CP-$odd Higgs boson ~$A^0$ ~\cite{BF},
which do not affect ~$\Delta\rho$ ~too much.
Therefore, in the MSSM the requirement of a good overall fit is not in conflict
with requirement of an enhancement of ~$R_b$ ~\cite{MY_MSSM} and they
imply a hierarchy:
\begin{eqnarray}
M_{\tilde t_L} >> M_{\tilde t_R}  ~~~{\rm or} ~~~
M_{\tilde t_2} >> M_{\tilde t_1}
\label{eqn:hier}
\end{eqnarray}
with small left-right mixing.

The chargino  - stop loop contribution to the 
$Z^0b\bar{b}$ vertex can be with stop coupled 
to ~$Z^0$ ~and with charginos coupled to ~$Z^0$. ~In both cases the lighter
the stop and chargino the larger is the positive contribution.   
Moreover, the ~$b\tilde t_1 C^-$ ~coupling is enhanced for a right handed stop
(it is then proportional to the top quark Yukawa coupling). Then, however,
the stop coupling to ~$Z^0$ ~is suppressed (it is proportional to ~
$g \sin^2\theta_W$) ~and significant contribution can only come from the
diagrams in which charginos are coupled to ~$Z^0$. ~
Their actual magnitude depend on the interplay of the couplings in the ~
$C^-_i\tilde t_1 b$ ~and the ~$Z^0C^-_iC^-_j$ ~
vertices. The first one is large only for charginos with large up-higgsino
component, the second - for charginos with large gaugino component in at least
one of its two-component spinors. It has been observed that, this combination 
never happens for ~
$\mu>0$. Large ~$R_b$ ~can then only be achieved at the expense of very
light ~$C^-_j$ ~and ~$\tilde t_1$. ~
In addition, for fixed ~$m_{C_1}$ ~and ~$M_{\tilde t_1}$, ~
$R_b$ ~is larger for ~$r\equiv M_2/|\mu|>1$ ~i.e. for higgsino-like 
chargino as the enhancement
of the ~$C^-_1\tilde t_1 b$ ~coupling is more important than of the ~
$Z^0C^-_1C^-_1$ ~coupling.

For ~$\mu<0$ ~the situation is different. In the range ~
$r\approx1\pm0.5$ ~a light  chargino can be a strongly mixed state 
with a large up-higgsino and gaugino components (the higgsino-gaugino mixing
comes from the chargino mass matrix). Large couplings 
in both vertices of the diagram with charginos coupled to ~
$Z^0$ ~give significant increase in ~$R_b$  ($\sim 0.218$ for $m_{\tilde t}=
50$ GeV) ~even for the 
lighter chargino as heavy as ~$80-90$ GeV ~(similar increase in ~$R_b$ ~for ~
$\mu>0$ and the same stop masses ~requires ~$m_{C_1}\approx50$ GeV). ~

Significant enhancement of ~$R_b$ ~is also possible for large ~$\tan\beta$ ~
values, ~$\tan\beta\approx m_t/m_b$ ~\cite{BF}.
In this case, in addition to the stop--chargino 
contribution there can be even larger positive contribution from
the ~$h^0$, ~$H^0$ ~and ~$A^0$ ~exchanges in the loops, provided those
particles are sufficiently light (in this range of ~$\tan\beta$, ~
$M_h\approx M_A$) ~and non-negligible sbottom--neutralino loop
contributions. ~The main difference with the low ~$\tan\beta$ ~
case is the independence of the
results on the sign of ~$\mu$ ~(which can be traced back to the approximate
symmetry of the chargino masses and mixings under ~$\mu\rightarrow -\mu$).

As stressed earlier an enhancement in ~$R_b$ ~
must be subject to constraints from the quality of the global fit to
the electroweak data and from all other available experimental information. 
Those constraints often differ in the degree of their model dependence and are 
worth careful discussion. A good quality of a global fit to the data is mainly 
assured by heavy enough left-handed sfermions with no direct impact on the 
value of ~$R_b$. ~The main remaining effect is the contribution of the 
decays  ~$Z^0\rightarrow N_i^0N_j^0$ ~to the total and hadronic
widths, ~$\Gamma_Z$ and $\Gamma_h$, respectively. ~
One may wonder how this contribution depends on the assumption about the 
neutralino masses and their composition. With the GUT assumption, ~
$M_1=\frac{5}{3} \tan^2\theta_W M_2$, ~the neutralino mass matrix is determined
by the chargino one but, in general, one may consider arbitrary values
of ~$M_1$.~ It is remarkable that for low $\tan\beta$ and for fixed
ratio ~$r$ ~and ~$M_2$ ~the ~
$\Delta\Gamma=\Sigma\Gamma_{Z\rightarrow N^0_i N^0_j}$ ~
is very weakly dependent on the value of ~$M_1$.  ~In Fig.1a we show
the contour ~$\Delta\Gamma =5$ MeV ~for several values of ~$M_1$ ~and ~
$\mu < 0$ ~(for $\mu > 0$ the ~$\Delta\Gamma$ ~
is much smaller and is not relevant for the quality of the fit).
The contributions of the individual channels change with a change of the
neutralino masses and compositions but their sum remains almost constant
when ~$M_1$ ~is changed arbitrarily.  On the other hand, we see in Fig.1a
that for fixed ~$m_C$ ~the dependence on the ratio $r$ is strong. For instance,
for ~$m_C=70$ GeV, ~$\Delta\Gamma <5$ MeV ~only for ~$r>3$ ~or ~$r<0.7$ ~

Similar information for large ~$\tan\beta$ ~is shown in Fig.1b. Here the 
results are symmetric with respect to the sign of ~$\mu$. ~The dependence
of ~$\Delta\Gamma$ ~on the ~$M_1$ ~is not negligible 
and (this time for both signes of ~$\mu$) ~the larger the ~$M_1$ ~the 
largerger the region in ~$(r,m_C)$ ~plane where ~$\Delta\Gamma < 5$ MeV.

The ~$Z^0$ ~decay width into neutralinos has important impact on the quality
of the global fit and constrains the region of light charginos.
Thus, it puts a limit on the realistic increase in ~$R_b$. ~
Fig.2 helps to understand this point in a clear way.
We depict there the bound ~$\Delta\Gamma<5$ MeV ~together with contours
of constant ~$\delta R^{SUSY}_b$. ~The full results for ~$R_b$ ~
as a function of ~
$m_C$ ~(obtained from a scan over the parameters ~$\theta_{\tilde t}^{LR}$, ~
$\alpha_s(M_Z)$ ~with ~$M_{\tilde t_1}=55$ GeV, ~
$M_{\tilde t_2}$, ~$M_{\tilde f}$ ~(masses of
other squarks and sleptons) fixed to 1 TeV; in the case of large ~
$\tan\beta$, ~$M_A$ ~was fixed to ~55 GeV ~and  ~$M_{\tilde b_R}=130$ GeV) ~
together with the corresponding values of the ~$\chi^2$ ~
are shown in Fig.3a and 3b, for ~$\mu <0$ ~and in Fig.4 for ~
$\mu > 0$. ~The observed rise in the ~$\chi^2$ ~for light charginos 
is precisely due to large values of ~$\Delta\Gamma$.

The results presented in Fig.3b and 4 are valid as long as we do not assume
the $R-$parity conservation
\footnote{The direct contribution to $R_b$ from the diagrams with 
the $R-$parity violating couplings 
is generically negligible \cite{MY_DEB} and is not included here. We merely
discuss the r\^ole of the assumption about $R-$parity conservation
in constraining the dominant MSSM contribution to $R_b$}.
In this  case the only additional constraints are the
model independent bounds\\
$m_{C^-}>47$ GeV,\\
$M_{\tilde t_1}> 46$ GeV,\\
$M_h>60$ GeV,\\
$M_A>55$ GeV ~(for large ~$\tan\beta $).\\
The first two bounds are implicitly included in Fig. 3 and 4.
The r\^ole of the lower limit on the Higgs boson mass (for a compact formula
for radiatively corrected lighter Higgs boson mass in the limit ~$M_A>>M_Z$ ~
see \cite{HEMPF}) depends on the mass of the heavier stop and  the 
left-right mixing angle. For ~$M_{\tilde t_2} > 500$ GeV ~(as required for
good quality of the global fit) and small mixing angles (necessary for
large ~$R_b$) ~$M_h$ ~is above the experimental limit 
\footnote{Important r\^ole of the experimental lower bound on ~$M_h$ ~
          in ref. \cite{ELONA} in constraining the potential increase of ~
          $R_b$ ~is due to the 
          chosen upper bound ~$M_{\tilde t_1} < 250$ GeV ~which, anyway,  
          looks too low from the point of view of a global fit.}
in a large range of the parameter space. Very small and large 
left-right mixing angles are, however, ruled out by this constraint
\cite{MY_RB}. The Higgs boson mass bound
has been imposed on the results of Fig. 3 and 4.

Assuming $R-$parity conservation, one has much stronger constraints which 
follow from the existence of the LSP and the corresponding decay signatures of
the superpartners. The most important one are:\\
$m_{C^-}>65$ GeV for ~$|m_{C^-}-m_{N^0}|>5$ GeV \cite{LEP1.5},\\
$\Gamma(Z^0\rightarrow N^0_1N^0_1)<4$ MeV ,\\ 
$BR(Z^0\rightarrow N^{0}_{1}N^{0}_{2})<10^{-4}$,\\
D0 exclusion plot in ~$(M_{\tilde t_1}, m_{N^0_1})$ ~ 
for $m_{C^-_1}>M_{\tilde t_1}$ ~\cite{D0_EXCL},\\
$M_2 > 36$ GeV ~from the gluino search under the asumption ~
$M_3\approx3M_2$ \cite{ROY}.\\ 

The bound on the ~
$BR(Z^0\rightarrow N^0_1 N^0_2)$ ~is also shown in Fig.1 and 2. For small ~
$\tan\beta$ and ~$\mu < 0$ ~this bound is stronger than the effect of ~
$\Delta\Gamma$ ~and puts a lower limit on ~$m_C$ ~which is often stronger
than the direct limit of ~65 GeV ~(see Fig.3c where 
the full results obtained under the assumption of $R-$parity conservation
are presented). The values of the ~$\chi^2$ ~as a function 
of the chargino mass (in the now allowed
range of ~$m_C$) ~are almost identical to Fig.3a.

For low ~$\tan\beta$ ~and ~$\mu > 0$ ~the 
constraint disappears and only the direct limit on ~$m_C$ ~is relevant.
For large ~$\tan\beta$ ~(and both signs of ~$\mu$) ~the bound is relevant only
for ~$r<1$ ~whereas the increase in ~$R_b$ ~comes mainly for ~$r>1$. ~
Therefore, the results shown in Fig.4
do not depend on whether $R-$parity is conserved or explicitly broken.

Light chargino and stop as well as light charged 
Higgs boson induce non-standard 
top quark decays ~($t\rightarrow\tilde t_1 N^0_i$
and ~$t\rightarrow H^+ b$). ~Present experimnetal limits on those
branching ratios are very uncertain. On the theoretical side, e.g.
the branching ratio for ~$t\rightarrow\tilde t_1 N^0_i$ ~in the low ~
$\tan\beta$ region depends strongly on the properties of the neutralinos
and, generically, is acceptable for a gaugino - like neutralino \cite{ROY}.
In Fig.5 we show contours of this branching ratio for ~$\tan\beta=1.4$ 
and ~$\mu<0$, ~with the other parameters fixed to
the values which give the results for ~$\delta R_b$ ~which are 
shown in Fig.2.

Finally, light chargino and stop play important role in the FCNC
transitions. This is discussed in more detail in the next section.
This last constraint is also included in the full results for ~$R_b$ ~
presented in Fig.3 and 4.

The overall summary of the results presented in Fig.3 and 4 is as
follows: Even with no assumption about the $R-$parity conservation
and for arbitrary ~$M_1$, ~the realistic upper bound
for ~$R_b$ ~in the MSSM is $\sim 0.218 ~(0.2185)$ ~for low (large) ~$\tan\beta$
values and for both signs of $\mu$. Further increase is bounded by too large
values of ~$\Delta\Gamma$. ~In this general case, 
and with maximal ~$R_b$, ~
the chargino mass is as low as ~50 GeV ~for small ~$\tan\beta$ ~and ~$\mu>0$ ~
but ~$m_C > 60$ GeV ~for ~
$\mu < 0$ ~and always for large ~$\tan\beta$. ~In the MSSM with $R-$parity
conservation, the increase in ~$R_b$ ~is even more strongly constrained
(mainly by the direct limit ~$m_C > 65$ GeV ~and by the ~$BR(Z\rightarrow
N^0_1 N^0_2)$) ~
and generically smaller by ~$\sim 0.001$, ~except for the region of
low ~$\tan\beta$ ~($\sim1-2$) ~and ~$\mu < 0$. ~
There, ~$R_b\sim 0.218$ ~is still
realistic for ~$m_C\sim (70-90)$ GeV. ~The right-handed stop
can be around ~50 GeV ~but even with ~$M_{\tilde t}\sim (60-70)$ GeV ~the
effect on ~$R_b$ ~is not negligible. An increase in ~$R_b$ ~gives a shift
in the fitted value of ~$\alpha_s(M_Z)$: ~$\delta\alpha_s=-4\delta R_b$. ~
Therefore, ~$\delta R_b\sim (0.001-0.002)$ ~gives ~
$\alpha_s(M_Z)\sim (0.118-0.114)$. ~Finally, it is important to observe
that the maximal enhancement in ~$R_b$ ~in the MSSM compared to the SM value is
only of the order of the ~$1.5\sigma$ ~present experimental error.
Significant improvement of the experimental precision is necessary
to confirm such a difference between the two theories.
\vskip 0.5cm

{\bf 3. Rare processes with light chargino and stop.}
\vskip 0.3cm

There are several well known supersymmetric contributions to rare processes.
In particular, supersymmetry may provide new sources of flavour violation
in the soft terms. However, even assuming the absence of such new effects,
there are obvious new contributions when  ~$W^{\pm}- q$ ~SM loops are 
replaced by the ~$H^{\pm} -q$ ~loops and by ~
${\tilde W^{\pm}} ({\tilde H^{\pm}}) - {\tilde q}$ ~loops. Those can be 
expected to be very important in the presence of light chargino and stop 
and they contribute to all best measured observables: ~$\epsilon-$ parameter
for the ~$\overline K^0 - K^0$ ~system, ~$\Delta m_B$ ~from ~
$\overline B^0-B^0$ ~mixing and ~$BR(b\rightarrow s\gamma)$. 

There are two important facts to be remembered about these contributions. 
They are present even if quark and squark mass matrices are diagonal in the
same super-Kobayashi-Maskawa basis. However the couplings in the vertex ~
$d_i\tilde u_j C^-$ ~depend on this assumption and 
can depart from the ~$K$-$M$ ~
parametrization if squark mass matrices have flavour-off diagonal entries
in the super-Kobayashi-Maskawa basis. Some of those entries are still totally
unconstrained and this is precisely the case for the (right) up squark
mass matrix which is relevant e.g. for the couplings ~$b\tilde t_R C^-$. ~
Still, sizeable suppression compared to the ~$K$-$M$ ~parametrization requires 
large flavour-off diagonal mass terms, of the order of the diagonal ones. To
remain on the conservative side we include the constraints from rare 
processes under the assumption of the ~$K$-$M$ ~parametrization of 
the chargino vertices. The r\^ole of 
the ~$b\rightarrow s\gamma$ ~constraint is discussed in detail in ref. 
\cite{MY_RB}.
The requirement of the acceptable $BR(b\rightarrow s \gamma)$ has been
imposed on the results shown in Fig.3 and 4.

The second important remark is that the element ~
$V_{td}\approx A\lambda^3(\rho - i \eta)$ ~
(in the Wolfenstein parametrization),
which is necessary for the calculation of the chargino and charged Higgs boson
loop contribution  to the ~$\epsilon_K$ ~
parameter and the ~$\overline B^0- B^0$ ~mixing,
is not directly measured. Its SM value can change after the inclusion of
new contributions. Thus the correct approach is the following one: take e.g.
\begin{eqnarray}
\Delta m_B\approx f^2_{B_d} B_{B_d} |V_{tb} V_{td}^*|^2 |\Delta|
\end{eqnarray}
where
\begin{eqnarray}
\Delta = \Delta_W +\Delta_{NEW}
\end{eqnarray}
is the sum of all box diagram contributions, ~$f_{B_d}$ and ~$B_{B_d}$ ~are
the ~$B^0$ ~meson decay constant and the vacuum saturation parameter. The ~
$CP$ ~violating parameter ~$\epsilon_K$ ~can also be  expressed in terms of ~
$\Delta$. ~Given ~$|V_{cb}|$, ~and ~$|V_{ub}/V_{cb}$ ~(known from the
the tree level processes i.e. almost unaffected by the supersymmetric
contributions) one can fit the parameters ~$\rho$, ~$\eta$ ~and ~$\Delta$ ~
to the experimental values of ~$\Delta m_{B_d}$ ~and ~$|\epsilon_K|$. ~
This way we find \cite{BRANCO,ZWIRNER} a model independent constraint
\begin{eqnarray}
{\Delta\over\Delta_W} < 3
\end{eqnarray}
for ~$\sqrt{f^2_{B_d} B_{B_d}}$ ~in the range ~$(160-240)$ GeV ~and ~
$B_K$ ~in the range ~$(0.6 - 0.9)$ GeV, ~which in the next step  can be used
to limit the allowed range of the stop and chargino masses and mixings.
The parameter space which is relevant for
an increase in ~$R_b$ ~gives large contribution to ~$\Delta$. ~It is still 
consistent with the bound (3)  but requires modified (compared 
to the SM) values of the  ~$CP-$violating phase ~$\delta$ ~($\eta, \rho$).
\vskip 0.5cm

{\bf 4. $R_b$ and exotic events.}

A single event ~$e^+e^-\gamma\gamma + {\rm missing} ~E_T$ ~has been reported
by Fermilab \cite{EEGG_EXP}.  
Preliminary results from the LEP1.5 run (after the upgrade of
energy to ~$\sqrt{s} = 130-136$ GeV) ~include peculiar four-jet events 
reported by ALEPH \cite{ALEPH4J}. 
Although statistics is too low to exclude fluctuations,
it is interesting to speculate if they can be explained by supersymmetry and
whether simultaneous explanation of these events and the ~$R_b$ ~anomaly  is 
possible. A detailed study of the Fermilab event 
in the supersymmetric extension of the SM is a subject of refs. 
\cite{EEGGET}. It is interesting to observe that the Fermilab event can be 
explained as a selectron pair production, with the supersymmetric spectrum
which is consistent with larger than in the SM values of ~$R_b$. The best 
description is obtained for ~$M_1\approx M_2$ ~but in a model with ~
$R-$parity conservation. This last fact should be stressed in view of
the following discussion of ALEPH events.

ALEPH 4-jet events have very peculiar gross features. On the kinematical
grounds they can be interpreted as a production of a pair of new particles ~
$X$ ~with ~$m_X\approx55$ GeV ~and a relatively large effective (after cuts)
production cross section ~$\sigma\approx3.7\pm1.7$ pb. ~Any interpretation 
of this new particle is strongly constrained by the decay signature: to a good
approximation no missing energy has been observed and most of the events
do not contain identified ~$b-$quark jets and no fast leptons in the final 
state. Those signatures of the events imply that any explanation within a ~
$R-$parity conserving MSSM is very difficult (for an explanation based on
the idea of a light gluino, ~$m_{\tilde g}\sim 1$ GeV ~see ref. \cite{FARRAR}).
Moreover, a large production cross section is not easy to accomodate ~
($R-$parity violation has little impact on the production cross section so
it can be reliably estimated in the MSSM). 
A sneutrino pair production seems
to be an acceptable possibility \cite{BGH} but its connection to the ~$R_b$ ~
anomaly is not obvious. Turning now to supersymmetric fermions, a neutralino
of a mass ~55 GeV ~has production cross section more than one order of 
magnitude below the reported value. Thus, we are left with a light chargino
as the most interesting candidate to explain ALEPH events. Indeed, the full
production cross section are typically large ~${\cal O}(10$pb) ~(see Fig.6).
Moreover, there is an interesting link with ~$R_b$ ~anomaly. 

The question which remains is whether chargino decay signatures can be 
consistent with ALEPH data. No missing energy rules out ~$R-$parity conserving
schemes. If ~$R-$parity is not conserved, additional terms in the
superpotential are allowed
\begin{eqnarray}
W &=& \frac{1}{2} \lambda_{ijk} L_i L_j E^c_k
  + \lambda^{\prime}_{ijk} L_i Q_j D^c_k\nonumber\\
  &+& \frac{1}{2} \lambda^{\prime\prime}_{ijk} U^c_i D^c_j D^c_k
\end{eqnarray}
where ~$\lambda_{ijk} = -\lambda_{jik}$ ~and ~
$\lambda^{\prime\prime}_{ijk} = -\lambda^{\prime\prime}_{ikj}$. ~
The first two terms violate lepton number conservation 
and the last one - baryon number conservation. 
Simultaneous presence of both types of terms can
lead to a rapid proton decay. However, only ~$\lambda$ ~and ~
$\lambda^{\prime}$ ~or only ~$\lambda^{\prime\prime}-$type 
couplings are allowed, of
course within the present experimental limits. The latter depend on the type
of the coupling but for several of them are relatively weak, particularly 
for the couplings involving the third family. 

If the chargino decayed through a lepton number violating coupling, there 
should be a hard lepton or missing momentum in the event. Thus this 
explanation looks unlikely. With baryon number violating couplings ~
$\lambda^{\prime\prime}$, ~the chargino may decay via either of two channels
\begin{eqnarray}
C^{\pm}\rightarrow {\tilde q_1}^* q_2 \rightarrow q_2 q_3 q_4 
\label{stopdecay}
\end{eqnarray}
where the squark is right handed and can be virtual, and
\begin{eqnarray}
C^{\pm}\rightarrow N^{0*} W^{\pm *} 
\rightarrow & \hspace*{-.5em} N^{0*} f_1 f_2 \nonumber \\
            & \hookrightarrow q_3 q_4 q_5 
\label{neutdecay}
\end{eqnarray}
with ~$N^0$ ~real or virtual. The actual decay pattern depends on the details
of the couplings and the values of the masses. 
It was shown in ref. \cite{MY_DEB} that, with most natural
assumptions, that right stop is the lightest squark and that the coupling ~
$\lambda^{\prime\prime}_{tds}$ ~is the largest one, 
the decay mode (6) is the more interesting one. Although
at the qualitative level 
we may  expect in this scenario too many jets and/or ~$b-$quark jets
in the final state, it happens that the effective 
4-jet cross section (after the experimental cuts)
with at most one b-quark contributing to the visible
energy is of the right order of magnitude (see Fig.7a).
Particularly attractive is the possibility (Fig.7b) ~
$m_{C^-} \gsim M_{\tilde t_R}\approx55$ GeV ~and with 
the mass difference such that the stop is real but produced almost at rest.
Then ALEPH events can be explained by \cite{MY_DEB}
\begin{eqnarray}
Z^0\rightarrow C^-C^+ \rightarrow 
(\tilde t_R \overline b)(\tilde{\overline t_R}  b)
\rightarrow (\overline d \overline s \overline b) (d s b)
\end{eqnarray}
with very slow  ~$b-$quarks and therefore escaping detection. With the 
present experimental resolution, a mass degeneracy ~
$m_{C^-}-M_{\tilde t_R}\lsim5-12$ GeV ~is sufficient for this scenario
\cite{MY_DEB}. 
This case is unique in the sense of avoiding b-jets as well as giving
the correct jet invariant mass distribution with about half
of the events falling into the range
\begin{equation}
103.3 {\rm GeV}<\Sigma m (=m_{ij}+m_{kl})<106.6 {\rm GeV}
\end{equation}

The results for the chargino decay via virtual stop shown in Fig.6a
generically give much broader distributions for this sum of the 
properly paired two jet invariant masses.

Neutralino could still be light but 
the decay ~$C^-\rightarrow N^0 f_1f_2$ ~is
suppressed due to kinematical reasons (due to multibody final states).
A link with the ~$R_b$ ~anomaly is clear. However, simultaneous 
supersymmetric explanation of the Fermilab (one) event and the ALEPH ~
$4-$jet events looks unlikely because of the  need for broken ~$R-$parity
in the latter case.
\vskip 0.5cm

{\bf 5. Conclusions.}
\vskip 0.3cm

Chargino and right handed stop are likely to be (in addition to neutralinos)
the lightest supersymmetric particles. Not only the masses in the range or
even below ~$M_Z$ ~are not excluded by any presently available experimental
data, they may be responsible for an increase in $R_b$ up to $\sim 0.218$.
Depending on whether ~$R-$parity is conserved or not, the Fermilab event
or ALEPH events may be explained simultaneously with larger 
than the SM values of ~$R_b$. ~
In particular, ALEPH events, if confirmed, may signal the discovery of
a chargino with ~$m_{C}\approx60$ GeV, ~a stop with ~$M_{\tilde t}\approx55$
GeV ~and broken ~$R-$parity.

\newpage

\begin{figwindow}[0,l,%
{\epsfig{bbllx=200pt,bblly=250pt,bburx=425pt,bbury=630pt,figure=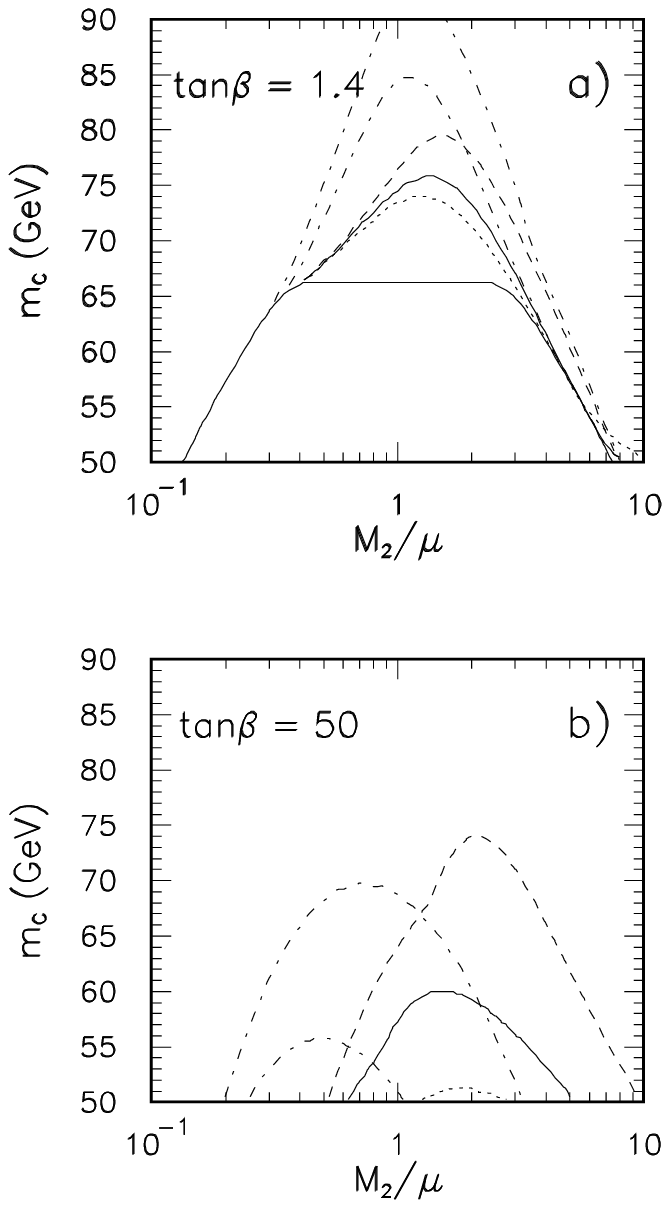,width=\textwidth,clip=}},%
{Dependence of the contour $\Delta\Gamma\equiv
\Sigma\Gamma_{Z\rightarrow N^0_i N^0_j}=5$ MeV on $M_1$. $M_1=0.2M_2$ (dashed),
$M_1=0.5M_2$ (solid) and $M_1=M_2$ (dotted). Dash-dotted lines
show $BR(Z\rightarrow N^0_1N^0_2)=10^{-4}$ for $M_1=0.2M_2$ (upper ones)
and $M_1=0.5M_2$ (lower ones). In a), the region below the lower solid line
is excluded by ``kinematics'' (the structure of the chargino mass matrix).}]
\end{figwindow} 

\newpage

\begin{figwindow}[0,l,%
{\epsfig{bbllx=85pt,bblly=230pt,bburx=540pt,bbury=630pt,figure=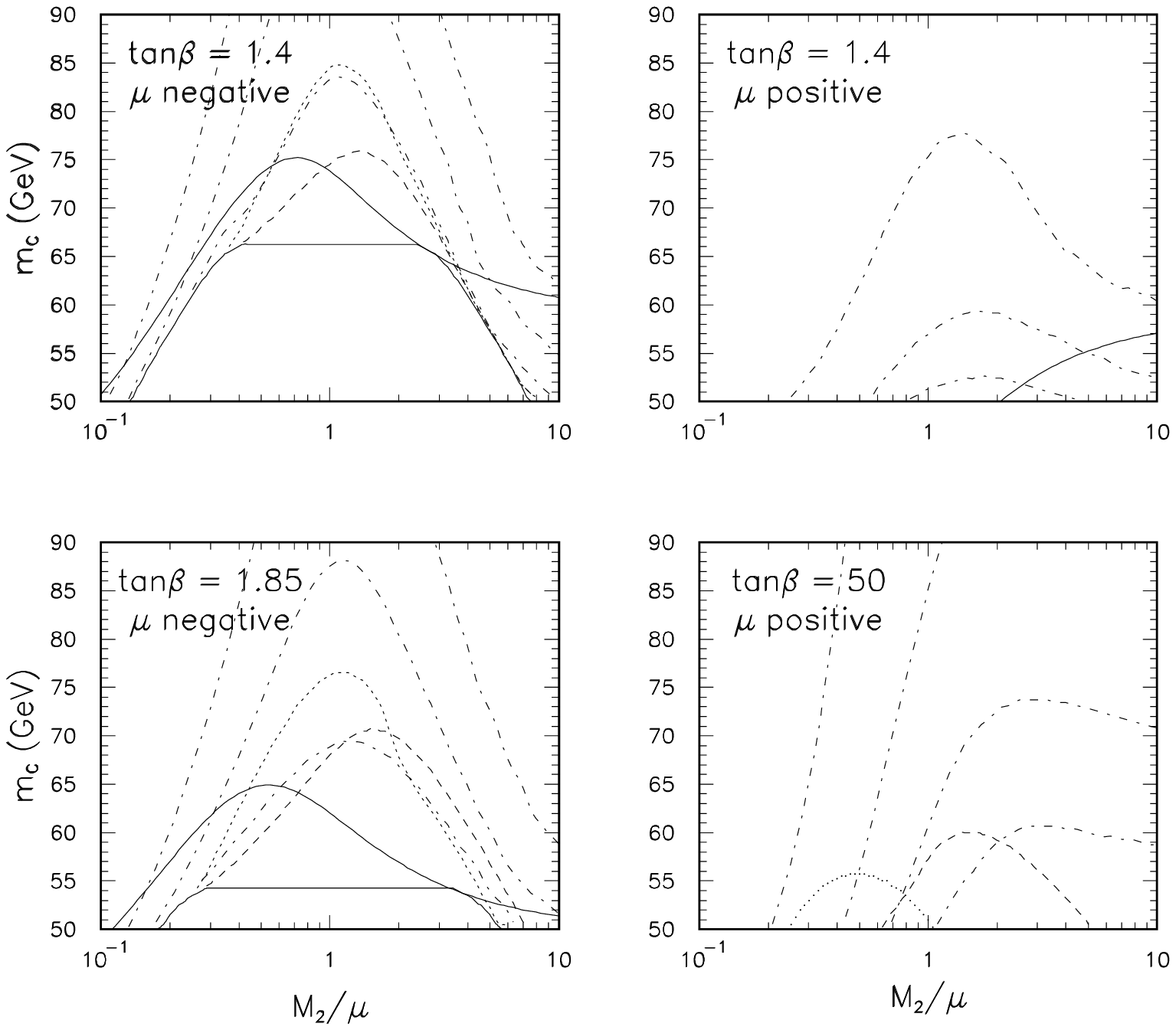,width=\textwidth,clip=}},%
{Contours of $\delta R^{SUSY}_b=$0.0020, 0.0015 and 0.0010 
(dash-dotted lines)
in the plane $(M_2/\mu, m_C)$ for $m_t=170$ GeV and different $\tan\beta$. 
Also shown are the lines of
$\Delta\Gamma=5$ MeV (dashed) and $BR(Z^0\rightarrow N_1N_2)=10^{-4}$
(dotted). Solid lines show the ``kinematic'' limits of $m_C$ and
$M_{\tilde g}=150$ GeV.}]
\end{figwindow} 

\newpage

\begin{figwindow}[0,l,%
{\epsfig{bbllx=60pt,bblly=155pt,bburx=540pt,bbury=715pt,figure=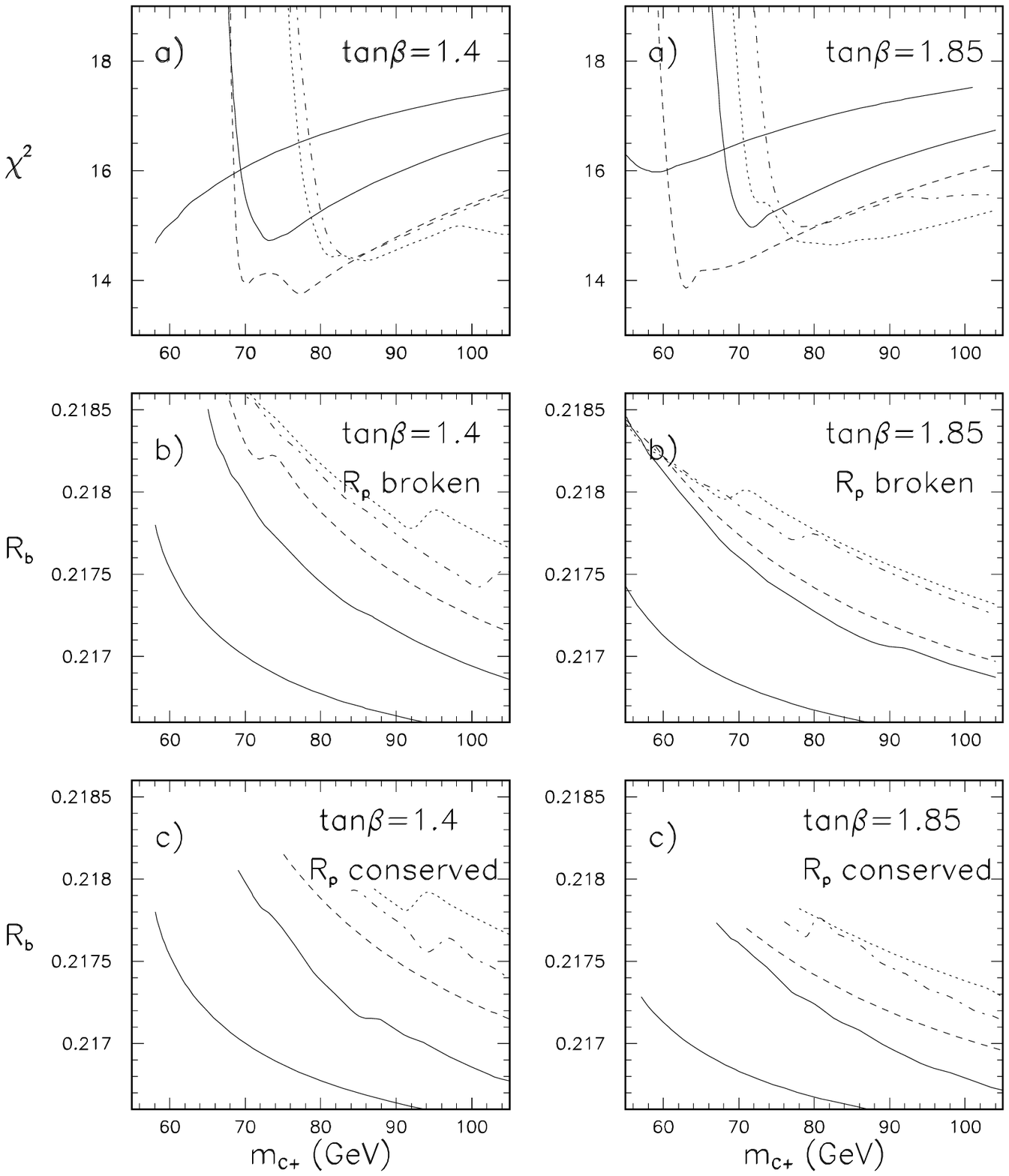,width=\textwidth,clip=}},%
{Best $\chi^2$ and the corresponding $R_b$ as a function of $m_C$, for
negative $\mu$, $m_t=170$ GeV, $M_{\tilde t_1}=50$ GeV,
$M_{\tilde t_2}=M_A=1$ TeV and for two values of $\tan\beta$.
For $R_b$ two cases are shown: with $R-$parity broken  and
with $R-$parity conserved. $\chi^2$ as a function of $m_C$ is
the same in both cases. Different lines correspond to different values
of $M_2/\mu$: 0.2 (upper (lower) solid for {\bf a} ({\bf b,c})),
0.5 (dashed), 1 (dotted), 1.5 (dash-dotted) and 3 (lower (upper) 
solid for {\bf a} ({\bf b,c})).}]
\end{figwindow} 

\newpage

\begin{figwindow}[0,l,%
{\epsfig{bbllx=80pt,bblly=230pt,bburx=540pt,bbury=635pt,figure=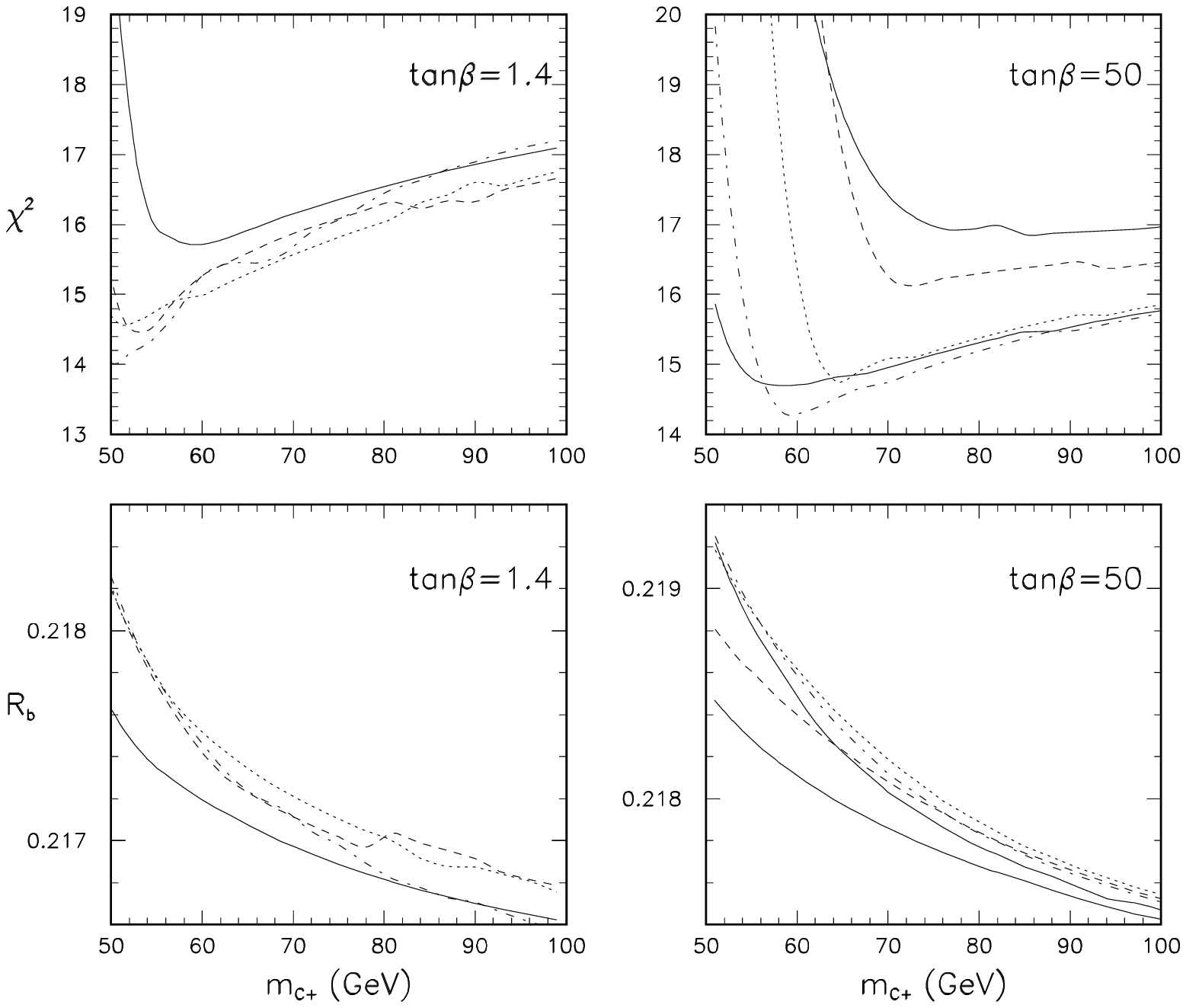,width=\textwidth,clip=}},%
{Best $\chi^2$ and the corresponding $R_b$ as a function of $m_C$, for
positive $\mu$, $m_t=170$ GeV, $M_{\tilde t_1}=50$ GeV,
$M_{\tilde t_2}=1$ TeV. $M_A=1$ TeV for $\tan\beta=1.4$ and
$M_A=55$ GeV for $\tan\beta=50$. Different lines correspond to different values
of $M_2/\mu$. For $\tan\beta=1.4$:
0.5 (solid), 1 (dashed), 1.5 (dotted) and 3 (dash-dotted);
for $\tan\beta=50$: 1 (upper (lower) solid for $\chi$ ($R-b$)),
1.5 (dashed), 3 (dotted), 5 (dash-dotted) and 10 (lower (upper) 
solid for $\chi$  ($R_b$)).
Identical results are obtained for broken or conserved $R-$parity.}]
\end{figwindow} 

\newpage

\begin{figwindow}[0,l,%
{\epsfig{bbllx=200pt,bblly=330pt,bburx=425pt,bbury=540pt,figure=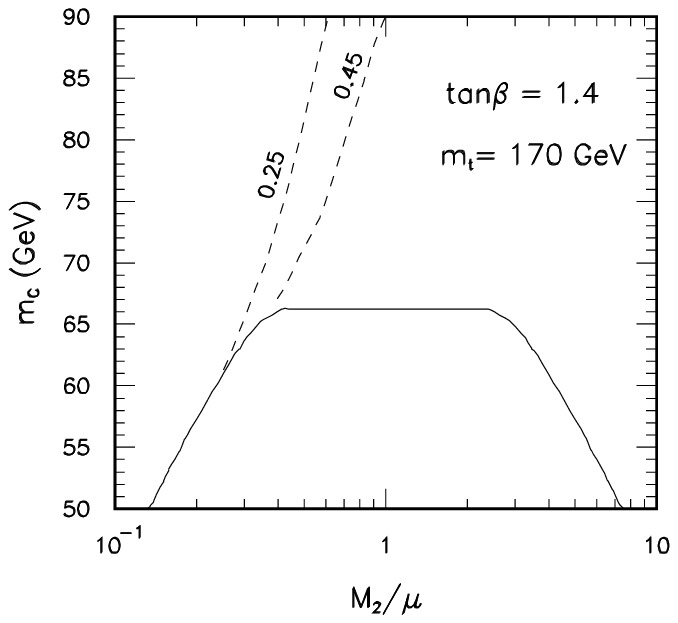,width=\textwidth,clip=}},%
{Contours of $BR(t\rightarrow\tilde t_1N^0_i)=$25\% 
and 45\% for ~$\tan\beta=1.4$, $\mu<0$ and the other parameters fixed at the 
values which give $\delta R_b$ as shown in Fig.2. The solid curve is the
``kinematical'' limit for the chargino mass.}]
\end{figwindow} 

\newpage

\begin{figwindow}[0,l,%
{\epsfig{bbllx=200pt,bblly=230pt,bburx=425pt,bbury=640pt,figure=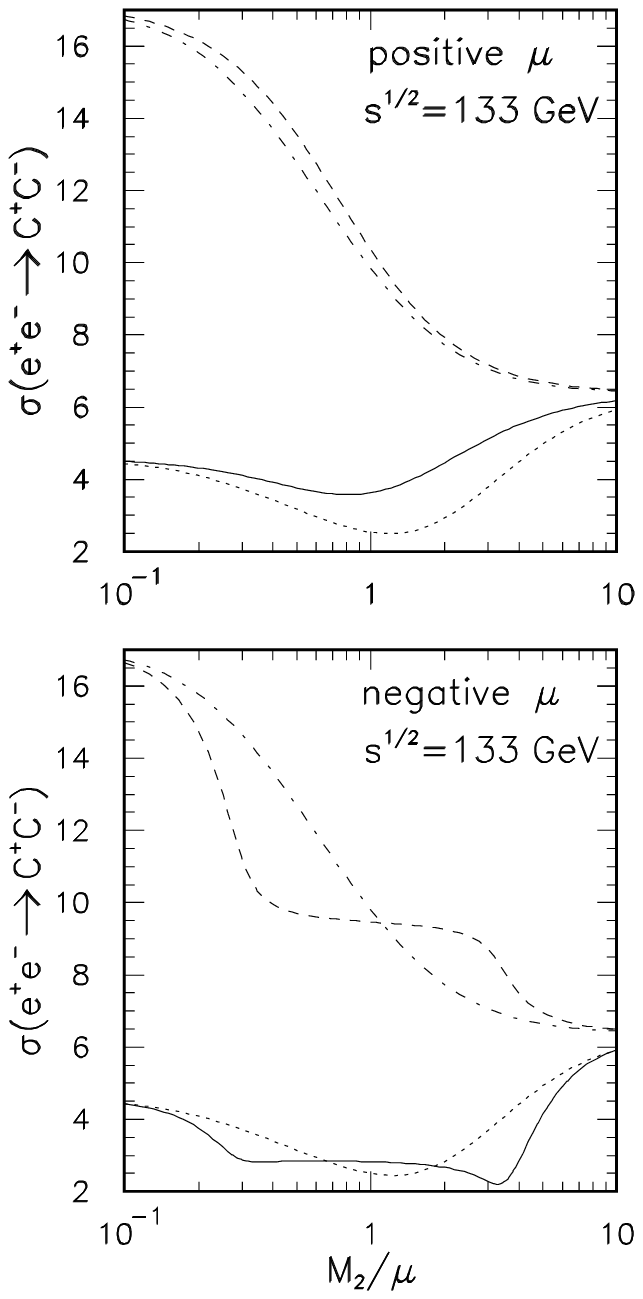,width=\textwidth,clip=}},%
{Cross sections for 55 GeV  chargino  production
for different choices of ~$(\tan\beta, M_{\tilde\nu_e})$ values:
for $\mu>0$: (1.4 50)-solid, (1.4,200)-dashed, (50,50)-dotted, 
(50,200)-dashdotted;
for $\mu<0$: (1.85, 50)-solid, (1.85,200)-dashed, (50,50)-dotted, 
(50,200)-dash-dotted.}]
\end{figwindow} 

\newpage

\begin{figure}[ht]
       \vskip 5in\relax\noindent\hskip -1.8in
       \relax{\includegraphics{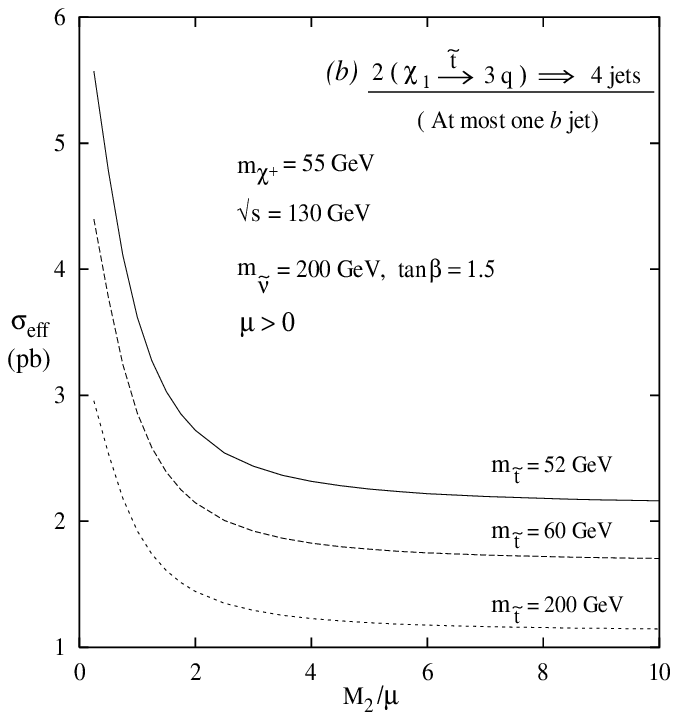}}
       \relax\noindent\hskip 3.25in
       \relax{\includegraphics{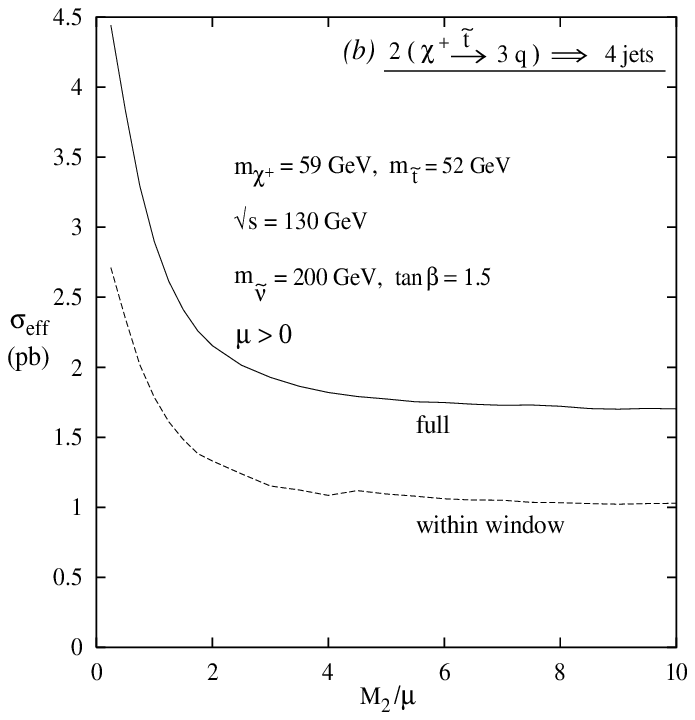}}
       \vspace{-20ex}
\caption{Effective 4-jet cross sections for pair-produced
charginos decaying through the stop channel: a) for 55 GeV chargino,
requiring that at most
one b-quark contributes to the visible energy, b) for 59 GeV chargino
decaying through real 52 GeV right stop (the lower (upper) curves
are with (without) the cut of eq.9).}
\label{fig:c_qs}
\end{figure}

\end{document}